\documentclass[aps,prl,reprint,superscriptaddress,showpacs]{revtex4-1}
\pdfoutput=1
\usepackage{color}
\usepackage{graphicx}
\usepackage{amsmath}
\usepackage{amsfonts}
\usepackage{amssymb}
\usepackage{rotating}
\usepackage{float}
\usepackage{verbatim}

\newcommand{\ket}[1]{|#1\rangle}
\newcommand{\braket}[2]{\langle #1|#2\rangle}
\newcommand{\Sd}[1]{S^{\dagger}_{#1}} 
\allowdisplaybreaks
\usepackage{multirow}
\usepackage{braket}
\usepackage{pdfpages}

\begin{document}

\author{Pieter W. Claeys}
\email{PieterW.Claeys@UGent.be}
\affiliation{Center for Molecular Modeling, Ghent University, Technologiepark 903, 9052 Zwijnaarde, Belgium}
\affiliation{Ghent University, Department of Physics and Astronomy, Proeftuinstraat 86, 9000 Ghent, Belgium}
\author{Stijn De Baerdemacker}
\affiliation{Center for Molecular Modeling, Ghent University, Technologiepark 903, 9052 Zwijnaarde, Belgium}
\affiliation{Ghent University, Department of Physics and Astronomy, Proeftuinstraat 86, 9000 Ghent, Belgium}
\affiliation{Ghent University, Department of Inorganic and Physical Chemistry, Krijgslaan 281 (S3), 9000 Ghent, Belgium}
\author{Dimitri Van Neck}
\affiliation{Center for Molecular Modeling, Ghent University, Technologiepark 903, 9052 Zwijnaarde, Belgium}
\affiliation{Ghent University, Department of Physics and Astronomy, Proeftuinstraat 86, 9000 Ghent, Belgium}

\title{Read-Green resonances in a topological superconductor coupled to a bath}

\begin{abstract}
We study a topological superconductor capable of exchanging particles with an environment. This additional interaction breaks particle-number symmetry and can be modelled by means of an integrable Hamiltonian, building on the class of Richardson-Gaudin pairing models. The isolated system supports zero-energy modes at a topological phase transition, which disappear when allowing for particle exchange with an environment. However, it is shown from the exact solution that these still play an important role in system-environment particle exchange, which can be observed through resonances in low-energy and -momentum level occupations. These fluctuations signal topologically protected Read-Green points and cannot be observed within traditional mean-field theory.
\end{abstract}

\pacs{74.20.Rp, 02.30.Ik,74.20.Fg,71.10.Li}
%74.20.Rp Pairing symmetries (other than s-wave)
%02.30.Ik Integrable systems
%74.20.Fg BCS theory and its development
%71.10.Li Excited states and pairing interactions in model system

\maketitle
\emph{Introduction.} --  Superconductivity is one of the most striking emergent features of many-body fermionic systems. The first successful description of this phenomenon was given by Bardeen, Cooper, and Schrieffer (BCS) by introducing a collective ground state consisting of condensed Cooper pairs and violating conservation of particle number \cite{bardeen_theory_1957}. This original mean-field theory was formulated for an $s$-wave pairing interaction, where the gap function is assumed isotropic. From the study of more general pairing interactions, it was later shown how topological superconductivity arose from a chiral $p_x+ip_y$-wave interaction by breaking time-reversal symmetry \cite{read_paired_2000,ryu_topological_2010}. Such pairing interactions are believed to occur naturally \cite{mackenzie_superconductivity_2003,xia_high_2006,maeno_evaluation_2011,kallin_chiral_2016} and have also been argued to be technologically achievable \cite{zhang_px+ipy_2008}. A major interest in these systems originates from their topological properties and the subsequent potential for quantum computation \cite{tewari_quantum_2007,sau_generic_2010,sarma_majorana_2015}.

Most of the theoretical insights into topological superconductivity are based on mean-field Bogoliubov-de Gennes theory, as initiated by Read and Green in their seminal paper \cite{read_paired_2000}. One of the crucial results was the uncovering of a phase transition between a (topologically nontrivial) weak-pairing and a (topologically trivial) strong-pairing state. At this transition, the chemical potential vanishes and the Bogoliubov quasiparticle spectrum becomes gapless. Alternatively, the theory of Richardson-Gaudin integrability also provides insights into topological superconductivity by means of the exact solution in finite-size systems \cite{dukelsky_colloquium:_2004}. The reduced $s$-wave pairing Hamiltonian for a finite system was solved exactly by Richardson in the 1960s \cite{richardson_restricted_1963,richardson_exact_1964,gaudin_bethe_2014}, but only rose to attention in the 1990s \cite{von_delft_spectroscopy_2001,amico_integrable_2001}. The exact solution was later generalized to $p_x+ip_y$ interactions through a variety of means \cite{ibanez_exactly_2009,skrypnyk_non-skew-symmetric_2009,dunning_exact_2010,rombouts_quantum_2010,VanRaemdonck2014,lukyanenko_boundaries_2014}, allowing for an exact calculation of spectral properties and correlation coefficients \cite{dunning_exact_2010,claeys_eigenvalue-based_2015}. From the exact solution, it was shown how the topological phase transition is reflected in the Read-Green points for finite systems. At these points, it is possible to reach excited states through a fixed number of zero-energy pair excitations. When a single zero-energy excitation is allowed, this corresponds to a vanishing chemical potential and the topological phase transition is recovered in the thermodynamic limit of the Richardson-Gaudin solution \cite{ibanez_exactly_2009,rombouts_quantum_2010}. This exact solution has also led to a criterion for the characterization of topological superconductivity in finite systems \cite{ortiz_many-body_2014,ortiz_what_2016}.

The main goal of this work is to put the robustness of the topological phase transition in a finite system to test when exchange of particles with an environment is allowed. For this, we consider a $p_x+ip_y$ superconductor coupled to a bath system by means of a recently-proposed integrable model, allowing for particle-number fluctuations \cite{lukyanenko_integrable_2016}. We present the exact Richardson-Gaudin eigenstates as well as exact correlation functions. As can be expected from a particle-number non-conserving Hamiltonian, the exact eigenstates mix separate $U(1)$-gauge (particle-number) symmetry sectors. Interestingly, they retain the factorized form of the Richardson-Gaudin particle-number conserving Hamiltonian. 

In this work, the exact solution is found to be essential to observe the effects of particle-exchange on the topological phase transition. It is shown how zero-energy excitations associated with the phase transition govern the particle-exchange with the bath, resulting in avoided crossings between states from different $U(1)$-symmetry sectors at the Read-Green points. These can be observed from strong fluctuations in the single-particle level occupations. This connects the physics of zero-energy excitations, arising from topological phase transitions, with the physics of open quantum systems. Furthermore, it is shown that while the breaking of particle-number symmetry in mean-field Bogoliubov-de Gennes theory has known great success in the description of number-conserving superconductors, it fails in describing the fluctuations observed in the exact solution.

\emph{The integrable model.} -- The particle-number conserving Hamiltonian of the integrable $p_x+ip_y$ (or $p+ip$) pairing model is given by \cite{ibanez_exactly_2009}
\begin{align}\label{hamnointeraction}
&H_{p+ip} = \sum_{\mathbf{k}}\frac{|\mathbf{k}|^2}{2m}c^{\dagger}_{\mathbf{k}}c_{\mathbf{k}}\nonumber\\
&\qquad-\frac{G}{4m}\sum_{\substack{\mathbf{k},\textbf{k}' \\ \mathbf{k} \neq \pm \textbf{k}'}}(k_x+ik_y)(k_x'-ik_y')c^{\dagger}_{\mathbf{k}}c^{\dagger}_{-\mathbf{k}}c_{-\mathbf{k}'}c_{\mathbf{k}'},
\end{align}
in which $c_{\mathbf{k}}$ and $c^{\dagger}_{\mathbf{k}}$ denote annihilation and creation operators respectively for two-dimensional spinless fermions of mass $m$ with momentum $\mathbf{k}=(k_x,k_y)$, and a dimensionless coupling constant $G$ has been introduced. The interaction with a bath can be modelled by introducing a coupling term
\begin{equation}\label{hamwithinteraction}
H =H_{p+ip}+\frac{\gamma}{\sqrt{2m}} \sum_{\mathbf{k}}\left[(k_x+ik_y)c^{\dagger}_{\mathbf{k}}c^{\dagger}_{-\mathbf{k}}+\text{h.c.}\right].
\end{equation}
This coupling term makes abstraction of the exact nature of the bath, and allows for particle exchange tunable by a single parameter $\gamma$. Alternatively, this additional term can be seen as a partial mean-field approximation of a more general $p_x+ip_y$ interaction Hamiltonian, allowing for particle-number fluctuations \footnote{Note that this interaction conserves fermion parity, so we will not consider the effect of Majorana zero modes associated with the spontaneous breaking of this symmetry.}. The BCS pairing gap of this model is then given by $\Delta=\frac{G}{m}\sum_{\mathbf{k}}(k_x+ik_y)\braket{c_{-\mathbf{k}}c_{\mathbf{k}}}-\frac{\gamma}{\sqrt{2m}}$, which is the pairing gap for an isolated system shifted by $\gamma/\sqrt{2m}$ \cite{ibanez_exactly_2009,rombouts_quantum_2010}.

Through the quasispin formalism \cite{talmi_simple_1993}, the Hamiltonian can be rewritten as  
\begin{align}
H=\sum_{k=1}^L \epsilon_{k}^2 \left( S_k^0+\frac{1}{2}\right)-G \sum_{k,k' \neq k}^L \epsilon_{k'} \epsilon_k S^{\dagger}_{k'}S_k \nonumber \\
+\gamma\sum_{k=1}^L \epsilon_k \left(\Sd{k}+S_k\right),
\end{align}
with $S^{\dagger}_{k}=\frac{k_x+ik_{y}}{\mathbf{k}}c^{\dagger}_{\mathbf{k}}c^{\dagger}_{-\mathbf{k}}$, $S_{k}$ its Hermitian conjugate, $S_k^0=\frac{1}{2}(c^{\dagger}_{\mathbf{k}}c_{\mathbf{k}} + c^{\dagger}_{-\mathbf{k}}c_{-\mathbf{k}}-1)$ and $\epsilon_k=|\mathbf{k}|/\sqrt{2m}$, where integers have been used to label the $L$ different states. The quasispin operators generate a set of quasispin-$1/2$ $su(2)$ algebras \footnote{The integrability only holds for doubly-degenerate levels, so we will restrict ourselves to spin-$1/2$. For the model at hand, this does not allow $|\mathbf{k}|=|\mathbf{k'}|$ if $\mathbf{k} \neq \pm \mathbf{k'}$.}. This Hamiltonian was recently shown to be integrable by Lukyanenko \emph{et al.} \cite{lukyanenko_integrable_2016} from the quasi-classical limit of the boundary quantum inverse scattering method, motivating our choice of bath.

\emph{Canonical eigenstates and Read-Green points.} -- We will first revisit the number-conserving Hamiltonian (\ref{hamnointeraction}), where an eigenstate containing $N$ fermion pairs (or $2N$ fermions) is given by a Bethe ansatz state
\begin{equation}\label{wavefunctionnointeraction}
\ket{\psi_N}=\prod_{\alpha=1}^N\left(\sum_{k=1}^L \frac{\epsilon_k }{\epsilon_k^2-v_{\alpha}^2}S^{\dagger}_k\right)\ket{\theta},
\end{equation}
with $\ket{\theta}$ denoting the particle vacuum state and the variables $\{v_{\alpha}^2,\alpha=1 \dots N\}$ (so-called rapidities) coupled through a set of Bethe ansatz equations (BAE) \cite{ibanez_exactly_2009}. The total energy of this state is (up to a constant) given by $(1+G)\sum_{\alpha=1}^N v_{\alpha}^2$, so each rapidity $v_{\alpha}^2$ can be loosely interpreted as the energy of a single Cooper pair. 

The BAE equations have the remarkable property that zero-energy fermion pair excitations are supported at specific fractional values of the coupling constant $G^{-1}=L-2N-p$, $p \in \mathbb{N}$, the so-called Read-Green points \cite{ibanez_exactly_2009,rombouts_quantum_2010,links_exact_2015}. Mathematically speaking, at $G^{-1}=L-2N-p$, if the set of $N$ rapidities $\{v_1^2,\dots, v_N^2\}$ is a solution to the BAE, then the set of $N+p$ rapidities $\{v_1^2,\dots, v_N^2,0,\dots,0\}$ is another solution to the BAE. The zero-solutions do not contribute to the energy, so the states defined by these variables, $\ket{\psi_N}$ and $\ket{\psi_{N+p}}$, are degenerate.

The $p=1$ case corresponds to a vanishing chemical potential in mean-field theory, since a fermion pair can then be added without changing the energy ($\ket{\psi_N}$ and $\ket{\psi_{N+1}}$ are degenerate). Indeed, it has been shown that in the thermodynamic limit a third-order topological phase transition occurs at this point, accompanied by nonanalytic behaviour of the ground-state energy \cite{rombouts_quantum_2010,ortiz_many-body_2014}. These zero-solutions can be contrasted to the gapless quasiparticles found in mean-field theory: while both correspond to zero-energy excitations, the latter is a single-fermion quasiparticle excitation, while the former corresponds to a collective fermion pair excitation, as can be seen from Eq. (\ref{wavefunctionnointeraction}). The value of the coupling constant $G^{-1}=L-2N-1$ where the phase transition occurs depends on the pair density $N/L$ but not on the single-particle energies $\epsilon_k$, since it is topologically protected. The integer topological invariant underlying this transition is given by a winding number denoting the transition between a topologically nontrivial and a topologically trivial state \cite{volovik_analog_1988,foster_quantum_2013}.

Note that both mentioned states are degenerate, but contain different numbers of fermions. When the Hamiltonian conserves particle number, these states are symmetry-protected and do not interact. However, once the symmetry is broken, e.g. by coupling to a bath, it's possible for these degenerate states to interact strongly, which will be illustrated after solving the Hamiltonian (\ref{hamwithinteraction}).

\emph{Solving the Bethe ansatz equations.} -- The main asset of integrable systems is that the diagonalization of a Hamiltonian matrix in an exponentially growing Hilbert space is reduced to solving a set of nonlinear equations scaling only linearly with system size. However, the BAE recovered for the extended Hamiltonian in the Supplemental Material are a great deal more involved than those for the canonical Hamiltonian, which are already notoriously difficult to solve generally. Instead of solving these equations directly, we will generalize a recent method pioneered by Faribault \emph{et al.} \cite{faribault_gaudin_2011,el_araby_bethe_2012,faribault_determinant_2012,tschirhart_algebraic_2014,faribault_determinant_2015} and later extended by us \cite{claeys_eigenvalue-based_2015,claeys_eigenvalue-based-bosonic_2015}. In this method, an algebraic relationship is obtained for the conserved operators in the integrable system, which are then converted to non-linear equations for their eigenvalues. These equations avoid the singularities plagueing the original BAE, and the rapidities can afterwards be efficiently extracted \cite{el_araby_bethe_2012}.

The set of conserved quantities associated to the integrable Hamiltonian are given by \cite{lukyanenko_integrable_2016},
\begin{align}\label{com}
&R_{k}=\left(S_k^0+\frac{1}{2}\right)-2G\sum_{k' \neq k}^L \frac{\epsilon_{k'}^2}{\epsilon_k^2-\epsilon_{k'}^2}\left(S_k^0S_{k'}^0-\frac{1}{4}\right) \nonumber\\
&+\gamma \epsilon_k^{-1} \left(S^{\dagger}_k + S_k \right) -G \sum_{k' \neq k}^L \frac{\epsilon_k \epsilon_{k'}}{\epsilon_k^2-\epsilon_{k'}^2}\left(S^{\dagger}_{k'} S_k+S_{k'} S^{\dagger}_k\right),
\end{align}
where $[H,R_{k}]=[R_{k},R_{k'}]=0$ and $H=\sum_k \epsilon_k^2 R_{k}$. From direct calculation it can be shown that the relations
\begin{equation}\label{evbeq}
R_{k}^2 = R_{k} +\gamma^2 \epsilon_k^{-2}+ G\sum_{k' \neq k}^L \epsilon_{k'}^2\frac{R_{k}-R_{k'}}{\epsilon_k^2-\epsilon_{k'}^2},\qquad \forall k,
\end{equation}
hold at the operator level. Since all operators commute mutually, all terms in this equation can be diagonalized simultaneously, and the eigenvalues $\{r_{k},\forall k\}$ can be found by solving this set of equations with conventional methods. 

\begin{figure}[htb!]                      
 \begin{center}
 \includegraphics[width=\columnwidth]{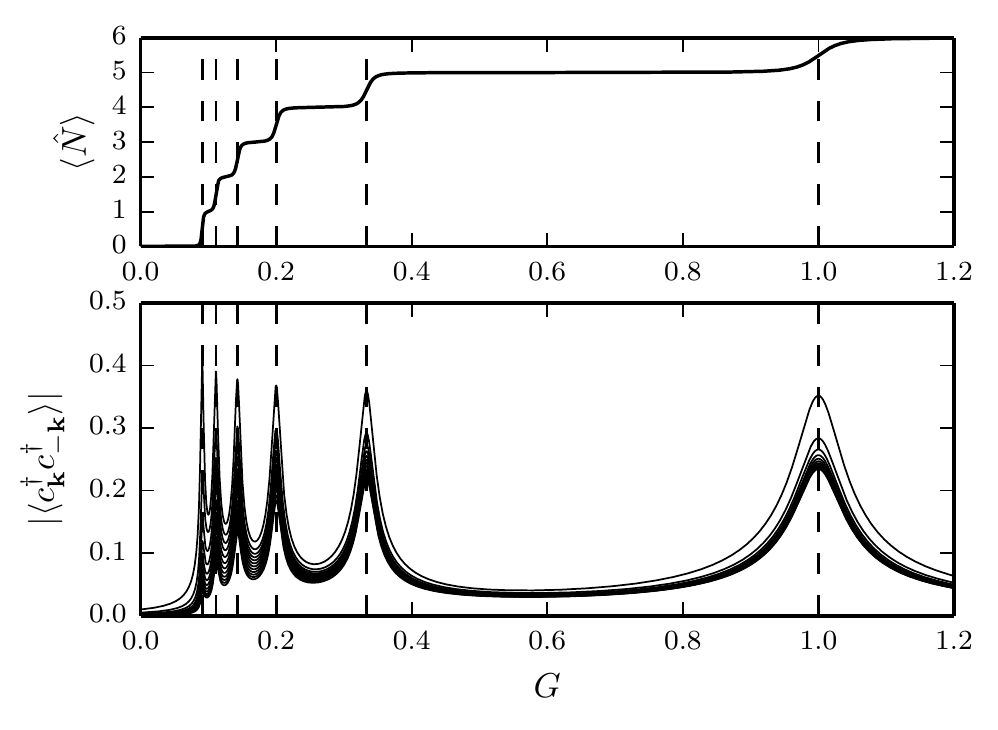}
 \caption{Average pair number $\braket{\hat{N}}$ and expectation values $\braket{c^{\dag}_{\mathbf{k}}c^{\dag}_{-\mathbf{k}}}$ for the ground state of  a picket-fence model ($L=12$) with $\gamma=10^{-2}$. At the Read-Green point $G^{-1}=L-2N-1$ the average pair number changes from $N$ to $N+1$, accompanied by sharp peaks in $\braket{c^{\dag}_{\mathbf{k}}c^{\dag}_{-\mathbf{k}}}, \forall \mathbf{k}$. \vspace{-\baselineskip}\label{groundstate}}
 \end{center}
\end{figure}

\emph{Signatures of the topological phase transition.} -- We first investigate signatures of the phase transition when a small interaction with a bath is introduced. Consider the population of the ground state $\braket{\hat{N}}$, with $\hat{N}$ the pair number operator, together with the expectation values $\braket{c^{\dag}_{\mathbf{k}}c^{\dag}_{-\mathbf{k}}}$ in Fig. \ref{groundstate}. It can be seen that the average number of Cooper pairs in the ground state increases by one unit at the crossing of a Read-Green point $G^{-1}=L-2N-1, N \in \mathbb{N}$. In other words, the systems absorbs a fermion pair from the environment at each Read-Green crossing. Note that the Read-Green points mark the phase transitions of isolated systems with different (fixed) densities, and the fluctuations in the density here lead to a series of Read-Green points, each associated with a different density.

\begin{figure}[htb!]                      
 \begin{center}
 \includegraphics[width=\columnwidth]{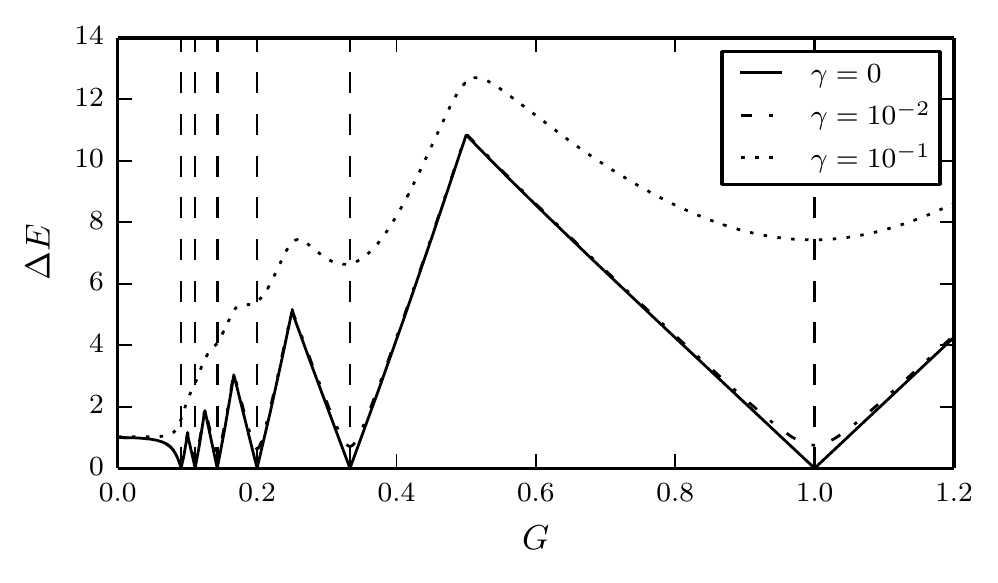}
 \caption{Energy gap $\Delta E$ between the first excited state and the ground state of a picket-fence model ($L=12$) for different values of $\gamma$. The Read-Green points $G^{-1}=L-2N-1$ with $N=0 \dots L/2-1$ are marked by vertical lines. \vspace{-\baselineskip}\label{gap}}
 \end{center}
\end{figure}

The mechanism underlying this particle exchange with the environment can be understood from Fig. \ref{gap}. At the Read-Green points, the isolated system ($\gamma=0$) is gapless, as follows from the definition of the Read-Green points, and a gap opens up for increasing $|\gamma|$. Where a phase transition is expected in the canonical regime with fixed pair density, a level crossing occurs instead when the exchange of particles with an environment is allowed. No transition from weak to strong pairing occurs, but the ground state jumps repeatedly from a weak pairing state $\ket{\psi_N}\ (N=0, \dots, L/2-1)$ to another weak pairing state with a higher pair density $\ket{\psi_{N+1}}$.

At the phase transition, the isolated system is gapless and the degenerate states are symmetry-protected due to the differing particle number. However, the interaction with the environment couples states with different particle number, and for small coupling $|\gamma|$, the states at the Read-Green points are (up to a perturbative correction) given by the coherent superposition $\frac{1}{\sqrt{2}}(\ket{\psi_N}-\ket{\psi_{N+1}})$, as follows immediately from perturbation theory. This strong deviation from the symmetry-protected states can be inferred from Fig. \ref{groundstate}, where the expectation values $\braket{c^{\dag}_{\mathbf{k}}c^{\dag}_{-\mathbf{k}}}$ are exactly zero when particle-number symmetry is conserved, but here exhibit sharp resonances exactly at the Read-Green points.

\begin{figure}[htb!]                      
 \begin{center}
 \includegraphics[width=\columnwidth]{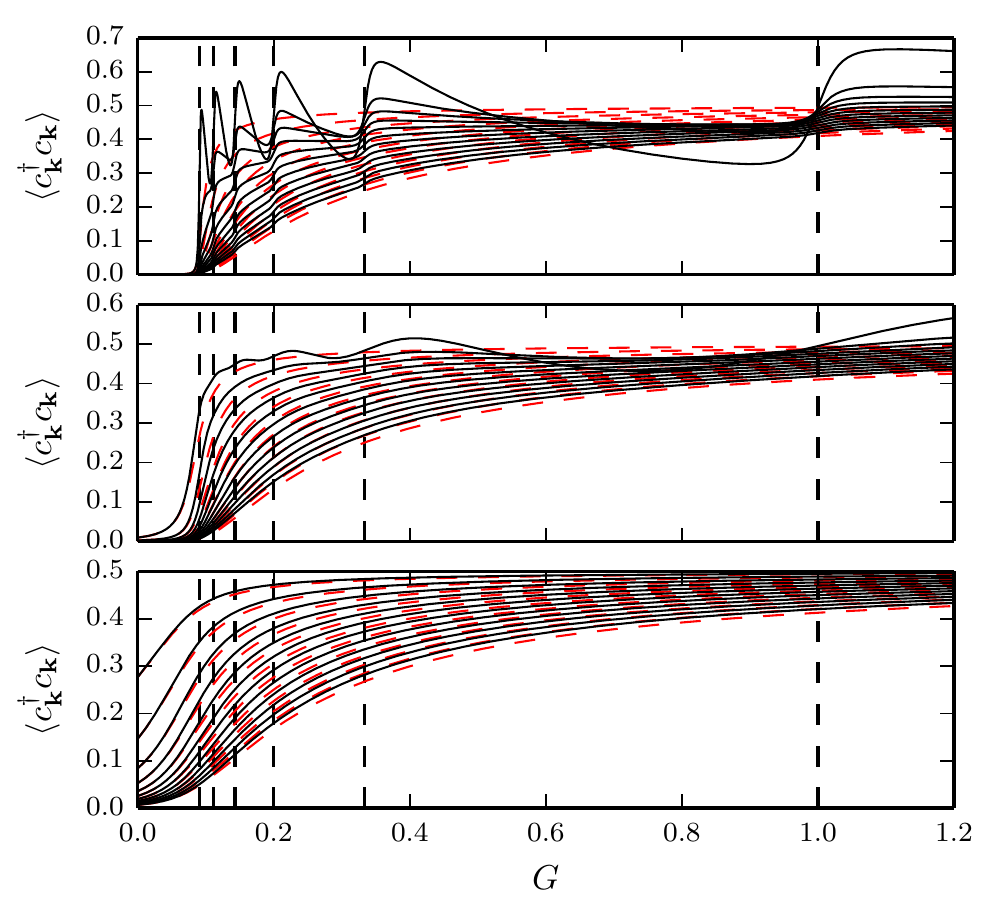}
 \caption{Occupation numbers of the ground state for varying $G$ at three different values of $\gamma=10^{-2},10^{-1}, 10^0$ (top to bottom). The exact solution is marked by black lines, while the mean-field solution is marked by red dashed lines. In the top figure, the lowest-momentum state exhibits peaks near the Read-Green points, which are again marked by vertical lines.\vspace{-\baselineskip}\label{expectationvalues}}
 \end{center}
\end{figure}

While the zero-energy excitations at the phase transition are not allowed for non-zero system-bath coupling, they can still be observed in the level occupations of the groundstate, as shown in Fig. \ref{expectationvalues}. For small $|\gamma|$, strong fluctuations are observed in the occupancy of the lowest-energy and -momentum states. These can be seen as the signatures of the zero-energy modes existing at each Read-Green point. For small interaction strengths the zero modes result in large fluctuations of the occupation of the lowest-energy and -momentum states, which exhibit sharp resonances near the Read-Green points, and can as such be termed \emph{Read-Green resonances}.

A comparison with the mean-field treatment on which the majority of theoretical insights for topological superconductors are based \cite{read_paired_2000} is also given in Fig. \ref{expectationvalues}, where we compare the exact distribution with the mean-field results. These mean-field results are based on the standard derivation for the $p_x+ip_y$-Hamiltonian \cite{read_paired_2000,botelho_quantum_2005,ibanez_exactly_2009,rombouts_quantum_2010}, where the chemical potential is set to zero and the gap instead contains a contribution from $\gamma$. For small $|\gamma|$ it can be seen that mean-field theory fails in capturing the fluctuations in the low-energy states, while for larger $|\gamma|$ the mean-field theory becomes increasingly more accurate. While it is known that mean-field methods are not adequate in finite-size systems, the extent to which they fail at detecting these resonances is remarkable. For large system-bath coupling the mean-field term in the Hamiltonian becomes dominant, so it is expected that mean-field theory will provide reliable results in this regime. From this, it is clear the regime with small system-bath coupling is the most physically interesting. As such, it should be noted that in order to be experimentally observable this regime requires a temperature significantly smaller than the pairing gap.

Although all calculations were performed on an integrable model, our results do not depend on the integrability of the interaction. They originate from the topological phase transition and the related zero-energy excitations, coupling states within different particle-number symmetry sectors. Apart from the ground state, Read-Green points are also spread throughout the entire spectrum, leading to avoided crossings and resonances within (highly) excited states. This connects these results with recent work on strong zero modes \cite{alicea_topological_2016}.

\emph{Rapidities.}  -- The structure of the eigenstates can also be used to shed light on the particle-exchange mechanism and its relation with zero-energy excitations. Following the methods of \cite{tschirhart_algebraic_2014}, the Hamiltonian (\ref{hamwithinteraction}) can be solved with a Bethe ansatz state
\begin{equation}\label{wavefunctionwithinteraction}
\ket{\psi^{\gamma}_L}=\prod_{\alpha=1}^L\left(\frac{\gamma}{ v_{\alpha}^2}+G\sum_{k=1}^L \frac{\epsilon_k \Sd{k}}{\epsilon_k^2-v_{\alpha}^2}\right)\ket{\theta},
\end{equation}
as a generalization of (\ref{wavefunctionnointeraction}) where the rapidities $\{v_{\alpha}^2,\alpha=1 \dots L\}$ are linked through the BAE presented in \cite{lukyanenko_integrable_2016} and derived in the Supplemental Material. This wavefunction again consists of a superposition of Cooper pairs, where each rapidity characterizes a single pair.

\begin{figure}[ht]                      
 \begin{center}
 \includegraphics[width=\columnwidth]{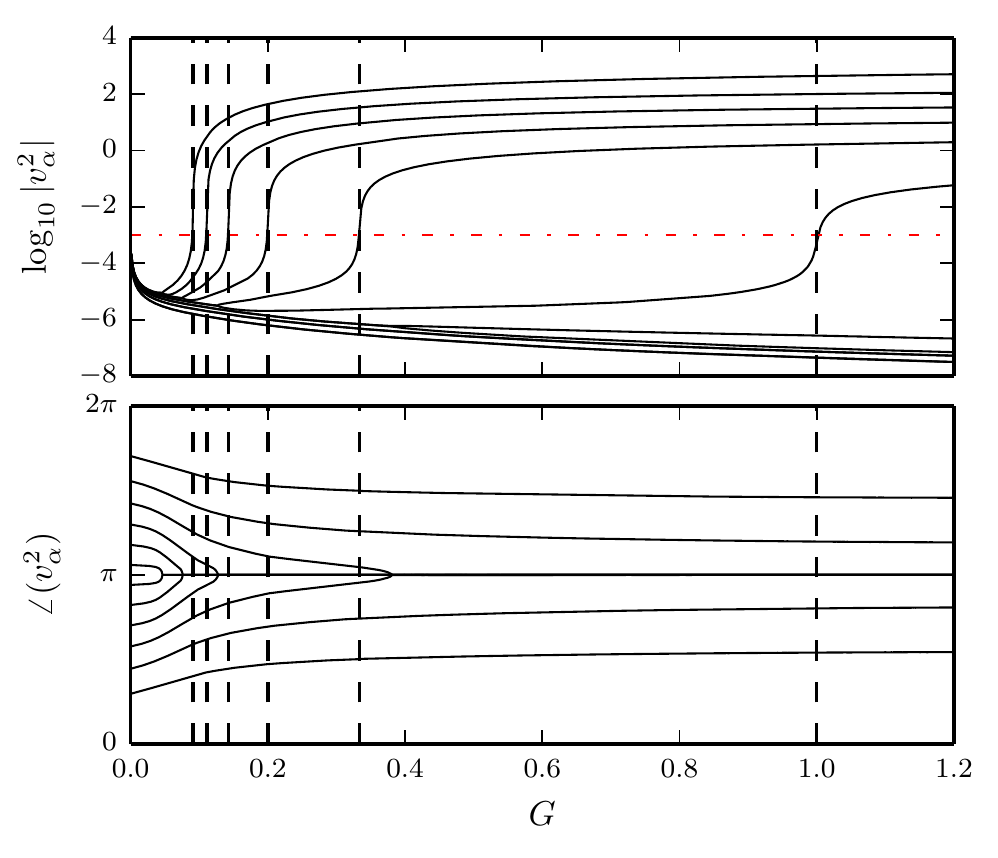}
 \caption{Modulus and phase ($\angle$) of the rapidities for the model in Fig. \ref{groundstate} with $\gamma=10^{-3}$ (marked by a dashed horizontal line). Note the logarithmic scale for the modulus. For decreasing $|\gamma|$ the transitions at the Read-Green points become steeper and the subset of rapidities below $|\gamma|$ decrease in magnitude, while the finite rapidities (above $|\gamma|$) remain approximately unchanged.\vspace{-\baselineskip}\label{rapidities}}
 \end{center}
\end{figure}

This parametrization allows for additional insight in the particle-exchange mechanism. From the factorized expression in Eq. (\ref{wavefunctionwithinteraction}) it can be seen that the factors for which $|v_{\alpha}^2| \ll |\gamma|$ only rescale the wavefunction (up to a small correction term) and do not lead to particle creation. Subsequently, if $L-N$ rapidities are small (compared to $|\gamma|$), the average number of Cooper pairs will be approximately $N$. Furthermore, the energy contribution of a single rapidity is proportional to $v_{\alpha}^{2}$, so these can be associated with zero modes. A clear separation of scales is seen in Fig. \ref{rapidities}, where at each Read-Green point a single rapidity quickly increases in magnitude, entailing a change by one in the average pair number. As such, the increase in average particle number reflects the activation of a single dormant zero-energy rapidity.

\emph{Conclusions.} -- In this work, it was shown how particle-exchange with an external environment influences the phase transition in a topological superconductor. At the topological phase transition, a single zero-energy excitation is created from the environment, increasing the average number of Cooper pairs in the superconductor, and the topological phase transition is changed to an avoided level crossing between topologically non-trivial states with different particle number. Each of these crossings is accompanied by a resonance in the level occupations of the lowest-energy single-particle states, which are a remainder of the zero-energy excitations and cannot be observed within traditional mean-field theory. This suggests identifying the topological Read-Green states by means of a coupling to an external bath.

\emph{Acknowledgements.} -- Pieter W. Claeys received a Ph.D. fellowship from the Research Foundation Flanders (FWO Vlaanderen).

\bibliography{MyLibrary.bib}
\end{document}

\author{Pieter W. Claeys}
\email{PieterW.Claeys@UGent.be}
\affiliation{Center for Molecular Modeling, Ghent University, Technologiepark 903, 9052 Zwijnaarde, Belgium}
\affiliation{Ghent University, Department of Physics and Astronomy, Proeftuinstraat 86, 9000 Ghent, Belgium}
\author{Stijn De Baerdemacker}
\affiliation{Center for Molecular Modeling, Ghent University, Technologiepark 903, 9052 Zwijnaarde, Belgium}
\affiliation{Ghent University, Department of Physics and Astronomy, Proeftuinstraat 86, 9000 Ghent, Belgium}
\affiliation{Ghent University, Department of Inorganic and Physical Chemistry, Krijgslaan 281 (S3), 9000 Ghent, Belgium}
\author{Dimitri Van Neck}
\affiliation{Center for Molecular Modeling, Ghent University, Technologiepark 903, 9052 Zwijnaarde, Belgium}
\affiliation{Ghent University, Department of Physics and Astronomy, Proeftuinstraat 86, 9000 Ghent, Belgium}

\title{Supplemental material}

%\pacs{74.20.Rp, 02.30.Rp,74.20.Fg,71.10.Li}
%74.20.Rp Pairing symmetries (other than s-wave)
%02.30.Ik Integrable systems
%74.20.Fg BCS theory and its development
%71.10.Li Excited states and pairing interactions in model system

\maketitle

In this supplemental material, we derive the Bethe ansatz equations (BAE) for the commuting operators $\{R_k\}$, which reduce to the conserved charges of the $p_x+ip_y$ model in the limiting case $\gamma=-\lambda$,
\begin{equation}
R_{k} =(S_k^0+\frac{1}{2})+\gamma \epsilon_k^{-1} S^{\dagger}_{k} -\lambda \epsilon_k^{-1} S_k - G \sum_{j \neq k}^L \left[\frac{\epsilon_k \epsilon_j}{\epsilon_k^2-\epsilon_j^2}\left(S^{\dagger}_{k} S_{j}+S_{k}S^{\dagger}_j\right)+\frac{2 \epsilon_j^2}{\epsilon_k^2-\epsilon_j^2}\left(S_{k}^0S_{j}^0-\frac{1}{4}\right)\right],
\end{equation}
starting from a product wave function
\begin{align}\label{pm:eigenstate}
\ket{\psi}=\prod_{\alpha=1}^L S^{\dagger}(v_{\alpha})\ket{\downarrow \cdots \downarrow}, \qquad S^{\dagger}(v_{\alpha})=-\frac{\lambda}{v_{\alpha}^2}+G\sum_{j=1}^L \frac{\epsilon_j}{\epsilon_j^2-v_{\alpha}^2} S^{\dagger}_{j},
\end{align}
with the operators $\{S_j^{\dagger},S_j,S_j^0,j=1 \dots L\}$ satisfying the commutation relations of a set of spin-$1/2$ $su(2)$ algebras and the set of $\{v_{\alpha}^2,\alpha=1 \dots L\}$ to be determined. This wavefunction does not contain a definite particle number due to the presence of the constant factor $\lambda/v_{\alpha}^2$ in the generalized creation operator $S^{\dagger}(v_{\alpha})$. Because of the commutativity, all conserved operators share a common set of eigenstates, and we will derive the conditions for the product state (\ref{pm:eigenstate}) to be an eigenstate of $R_{k}$. The action on the wave function can be determined through a Richardson-Gaudin commutator scheme as
\begin{align}
R_{k}\prod_{\alpha=1}^L S^{\dagger}(v_{\alpha})\ket{\downarrow \cdots \downarrow}=&\sum_{\alpha=1}^L\sum_{\beta=\alpha+1}^L\left(\prod_{\gamma \neq \alpha,\beta}^LS^{\dagger}(v_{\gamma})\right)[[R_{k},\Sd{\alpha}],\Sd{\beta}]\ket{\downarrow \cdots \downarrow} \nonumber\\
&+\sum_{\alpha=1}^L\left(\prod_{\beta \neq \alpha}^LS^{\dagger}(v_{\beta})\right)[R_{k},\Sd{\alpha}]\ket{\downarrow \cdots \downarrow}+\left(\prod_{\alpha=1}^L S^{\dagger}(v_{\alpha})\right)R_{k}\ket{\downarrow \cdots \downarrow}.
\end{align}
The necessary commutation relations are given by
\begin{align}
[R_k, S^{\dagger}(v_{\alpha})]&=-2 G \frac{v_{\alpha}^2}{\epsilon_k^2-v_{\alpha}^2}S^{\dagger}(v_{\alpha})S_k^0+2G^2 \frac{\epsilon_k}{\epsilon_k^2-v_{\alpha}^2}S^{\dagger}_k\left(\sum_{j=1}^L \frac{v_{\alpha}^2}{\epsilon_j^2-v_{\alpha}^2}S_j^0\right)+G\frac{\epsilon_k}{\epsilon_k^2-v_{\alpha}^2}S^{\dagger}_k\left(1+2G\sum_{j \neq k}^L S_j^0\right), \nonumber\\
[[R_k, S^{\dagger}(v_{\alpha})],S^{\dagger}(v_{\beta})]&=2G^2\frac{ \epsilon_k S^{\dagger}_k}{v_{\alpha}^2-v_{\beta}^2}\left[\frac{v_{\alpha}^2 }{\epsilon_k^2-v_{\beta}^2}S^{\dagger}(v_{\alpha})-\frac{v_{\beta}^2}{\epsilon_k^2-v_{\alpha}^2}S^{\dagger}(v_{\beta})\right].
\end{align}
Taking these results together, the action of a single conserved operator on the product wave function can be written as
\begin{align}
R_k \prod_{\alpha=1}^LS^{\dagger}(v_{\alpha})\ket{\downarrow \cdots \downarrow} =& \left[G\sum_{\alpha=1}^L \frac{v_{\alpha}^2}{\epsilon_k^2-v_{\alpha^2}}\right]\prod_{\alpha=1}^LS^{\dagger}(v_{\alpha})\ket{\downarrow \cdots \downarrow} +\gamma \epsilon_k^{-1}S^{\dagger}_k \prod_{\alpha=1}^LS^{\dagger}(v_{\alpha})\ket{\downarrow \cdots \downarrow}\nonumber\\
&+G\sum_{\alpha=1}^L \frac{\epsilon_k}{\epsilon_k^2-v_{\alpha}^2}\left[(1+G)-G\sum_{j=1}^L \frac{\epsilon_j^2}{\epsilon_j^2-v_{\alpha}^2}+2G\sum_{\beta \neq \alpha}^L\frac{v_{\beta}^2}{v_{\beta}^2-v_{\alpha}^2}\right]S^{\dagger}_k\prod_{\beta \neq \alpha}^L\Sd{\beta}\ket{\downarrow \cdots \downarrow}.
\end{align}
In order to be an eigenstate, the two unwanted (non-diagonal) terms should cancel. It can be shown that the contributions from different excitation sectors cancel termwise provided the variables $\{v_{\alpha}^2,\alpha=1 \dots L\}$ satisfy a set of Bethe ansatz equations. We first check the contribution from both terms to the state $S^{\dagger}_k\ket{\downarrow \cdots \downarrow}$ containing one excitation, which vanishes provided the set of equations
\begin{equation}\label{pm:matvecmult}
\sum_{\alpha=1}^L \frac{\epsilon_k^2v_{\alpha}^2}{\epsilon_k^2-v_{\alpha}^2}\left[(1+G)-G\sum_{j=1}^L \frac{\epsilon_j^2}{\epsilon_j^2-v_{\alpha}^2}+2G\sum_{\beta \neq \alpha}^L\frac{v_{\beta}^2}{v_{\beta}^2-v_{\alpha}^2}\right] = \frac{\gamma \lambda}{G}, 
\end{equation}
are satisfied. These can be brought in the previously-obtained form \cite{lukyanenko_integrable_2016}
\begin{equation}\label{pm:bae}
(1+G)-G\sum_{j=1}^L \frac{\epsilon_j^2}{\epsilon_j^2-v_{\alpha}^2}+2G\sum_{\beta \neq \alpha}^L\frac{v_{\beta}^2}{v_{\beta}^2-v_{\alpha}^2} = \frac{\gamma \lambda}{G} \frac{\prod_{j=1}^L(v_{\alpha}^{-2}-\epsilon_j^{-2})}{\prod_{\beta \neq \alpha}^L (v_{\alpha}^{-2}-v_{\beta}^{-2})},
\end{equation}
by interpreting (\ref{pm:matvecmult}) as a matrix-vector multiplication and multiplying with the well-known inverse of a Cauchy matrix. This requirement only cancels one contribution from both unwanted terms, all other contributions should also cancel exactly in order for (\ref{pm:eigenstate}) to be an eigenstate. The coefficient in front of a state with $N+1$ excitations where the set of spins labelled $\{i_1, \dots, i_N,k\}$ are flipped up is proportional to

\begin{align}\label{pm:unwantedterms}
C^N_{\{i_1, \dots, i_N,k\}}=\sum_{\alpha=1}^L \frac{F_{\alpha}}{\epsilon_k^{-2}-v_{\alpha}^{-2}}&\left[\sum_{\hat{A}\in S^N_{\hat{\alpha}}}\frac{1}{\prod_{j=1}^N(\epsilon_{i_j}^{-2}-\hat{\nu}^{-2}_{j})}\right]+\frac{\gamma\lambda}{G} \left[\sum_{A \in S^N}\frac{1}{\prod_{j=1}^N(\epsilon_{i_j}^{-2}-\nu_j^{-2})}\right],
\end{align}
with
\begin{equation}
F_{\alpha}=(1+G)-G\sum_{j=1}^L \frac{\epsilon_j^2}{\epsilon_j^2-v_{\alpha}^2}+2G\sum_{\beta \neq \alpha}^L\frac{v_{\beta}^2}{v_{\beta}^2-v_{\alpha}^2}
\end{equation}
and $S^N_{\hat{\alpha}}$ the set of all $N$-tuples built out of $N$ non-repeated elements $\{v_1, \dots v_{\alpha-1},v_{\alpha+1}, \dots, v_L\}$ and $S_N$ the set of all $N$-tuples built from $\{v_1, \dots, v_L\}$. The elements of these sets are denoted $\hat{A}_j=\{\hat{\nu}_1, \dots, \hat{\nu}_{N}\}$ and $A_j=\{\nu_1, \dots, \nu_N\}$. A highly similar expression was obtained in the study of the DJCG models, where it was proven that these expressions vanish provided the BAE equations are satisfied \cite{tschirhart_algebraic_2014}. The full proof that (\ref{pm:unwantedterms}) equals zero is completely analogous and does not depend on the explicit form of the BAE, rather on the structure from (\ref{pm:matvecmult}). 

\bibliography{MyLibrary.bib}